\documentclass[twocolumn,showpacs,preprintnumbers,amsmath,amssymb]{revtex4}
\usepackage{graphicx}
\usepackage{dcolumn}
\usepackage{bm}
\begin{document}
\title{Filling factor dependence of the fractional quantum Hall effect gap}
\author{V.~S. Khrapai, A.~A. Shashkin, M.~G. Trokina, and V.~T. Dolgopolov}
\affiliation{Institute of Solid State Physics, Chernogolovka, Moscow
District 142432, Russia}
\author{V. Pellegrini and F. Beltram}
\affiliation{NEST INFM-CNR, Scuola Normale Superiore, Piazza dei
Cavalieri 7, I-56126 Pisa, Italy}
\author{G. Biasiol and L. Sorba$^*$}
\affiliation{Laboratorio Nazionale TASC-INFM and NEST INFM-CNR,
I-34012 Trieste, Italy\\
$^*$Scuola Normale Superiore, Piazza dei Cavalieri 7, I-56126 Pisa,
Italy}
\begin{abstract}
We directly measure the chemical potential jump in the
low-temperature limit when the filling factor traverses the $\nu = 1/3$
and $\nu = 2/5$ fractional gaps in two-dimensional (2D) electron
system in GaAs/AlGaAs single heterojunctions. In high magnetic
fields $B$, both gaps are linear functions of $B$ with slopes
proportional to the inverse fraction denominator, $1/q$. The fractional
gaps close partially when the Fermi level lies outside. An empirical
analysis indicates that the chemical potential jump for an {\em ideal}
2D electron system, in the highest accessible magnetic fields, is
proportional to $q^{-1}B^{1/2}$.
\end{abstract}
\maketitle

The fractional quantum Hall effect in two-dimensional (2D) electron
systems \cite{tsui82} is believed to be a many-body phenomenon. The
dissipationless state manifested by zeros in the longitudinal
resistance and plateaus in the Hall resistance forms at fractional
filling factors of Landau levels, $\nu=p/q$, and the gap is predicted
to be caused by electron-electron interactions (see
Ref.~\cite{chakraborty00} and references therein). According to
numerical calculations, the chemical potential discontinuity
corresponding to the fractional gap is determined by $e^2/\varepsilon
l_B$ (where $\varepsilon$ is the dielectric constant and $l_B=(\hbar
c/eB)^{1/2}$ is the magnetic length), being almost independent of the
fraction, or the denominator $q$, for $1/3\le\nu\le2/3$
\cite{ambrumenil89,gros90,halperin93}. For quasiparticles with
fractional charge $e/q$ \cite{laughlin83}, the activation energy is
$q$ times smaller than the chemical potential discontinuity. This
determines the hierarchy of fractions, as inferred from the concept
of composite fermions \cite{jain89}. Notably, earlier studies predicted
a reduction of the chemical potential discontinuity with increasing $q$
for the same sequence \cite{halperin84}.

The fractional gap being small, its experimental determination is
pretty demanding with respect to both samples and methods. Attempts
to experimentally estimate the fractional gap value yielded similar
results in high magnetic fields
\cite{boebinger85,haug06,eisenstein92,dorozhkin93,hirji03,hirji05}
and, therefore, it is unlikely that the gap is strongly influenced by
the residual disorder in the 2D electron system. Standard
measurements of activation energy at the longitudinal resistance
minima were used to probe the fractional gaps with different
denominators \cite{boebinger85,du93,schulze04}. However, transport
studies yield a mobility gap which may be different from the gap in
the spectrum. The latter can be determined using direct thermodynamic
measurements of the chemical potential jump across the gap, similar
to those of Refs.~\cite{eisenstein92,khrapai07}. Recent thermodynamic
studies showed that in the completely spin-polarized regime, the gap
value for $\nu=1/3$ is equal to that for $\nu=2/3$, reflecting the
electron-hole symmetry in the spin-split Landau level, and increases
linearly with magnetic field $B$ \cite{khrapai07}. The experimental
gap behavior is in contradiction to the expected square-root
dependence of the gap on magnetic field, which reveals problems with
the straightforward consideration of electron-electron interactions
in 2D electron systems.

In this paper, we perform measurements of the chemical potential jump
across the fractional gap at filling factor $\nu=1/3$ and $\nu=2/5$
in the 2D electron system in GaAs/AlGaAs single heterojunctions using
a magnetocapacitance technique. The gap, $\Delta\mu$, increases with
decreasing temperature and saturates becoming independent of
temperature in the limit of low temperatures. In high magnetic
fields, the limiting gap values, $\Delta\mu_0$, for $\nu=1/3$ and
$\nu=2/5$ increase linearly with magnetic field so that the ratio of
the slopes is equal, within the experimental uncertainty, to the
inverse ratio of the fraction denominators. The
temperature-independent difference between the $\mu$ values at
$\nu=1/2$ and $\nu=1/4$ indicates that the fractional gap closes
partially when the Fermi level lies outside the gap. Unlike the spin gap
\cite{dolgopolov97}, the fractional gap decays due to temperature and
disorder smearing. Using an empirical analysis of the data, we allow for
the effect of sample inhomogeneities and extract the jump
$\Delta\mu^{id}(B)$ for an ideal 2D electron system. The results obtained
suggest that in the highest accessible magnetic fields, the gap
$\Delta\mu^{id}$ is proportional to $q^{-1}B^{1/2}$. We reach the limit of
very high magnetic fields where the expected square-root behavior
prevails (in homogeneous samples), while discrepancies between
experiment and theory may remain for lower $B$.

Measurements were made in an Oxford dilution refrigerator with a base
temperature of $\approx30$~mK on remotely doped GaAs/AlGaAs single
heterojunctions (with a low-temperature mobility $\approx4\times
10^6$~cm$^2$/Vs at electron density $9\times 10^{10}$~cm$^{-2}$)
having the quasi-Corbino geometry with area $2.2\times
10^5$~$\mu$m$^2$. The depth of the 2D electron layer was 210~nm. A
metallic gate was deposited onto the surface of the sample, which
allowed variation of the electron density by applying a dc bias
between the gate and the 2D electrons. The gate voltage was modulated
with a small ac voltage of 2~mV at frequencies in the range
0.05--11~Hz, and both the imaginary and real components of the
current were measured with high precision ($\sim10^{-16}$~A) using a
current-voltage converter and a lock-in amplifier. Smallness of the
real current component as well as proportionality of the imaginary
current component to the excitation frequency ensure that we reach
the low-frequency limit and the measured magnetocapacitance is not
distorted by lateral transport effects. A dip in the
magnetocapacitance in the quantum Hall state is directly related to
a jump of the chemical potential across the corresponding gap in the
spectrum of the 2D electron system \cite{smith85}:
\begin{equation}\frac{1}{C}=\frac{1}{C_0}+\frac{1}{Ae^2dn_s/d\mu},\label{C}\end{equation}
where $C_0$ is the geometric capacitance between the gate and the 2D
electrons, $A$ is the sample area, and the derivative $dn_s/d\mu$ of
the electron density over the chemical potential is the thermodynamic
density of states.

\begin{figure}
\scalebox{0.4}{\includegraphics[clip]{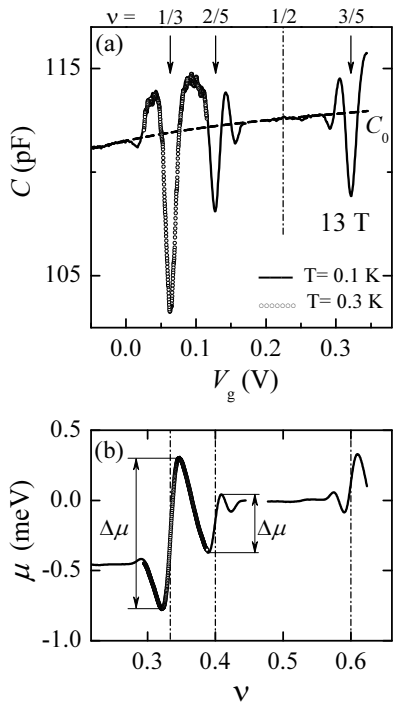}}
\caption{\label{fig1} (a)~Magnetocapacitance as a function of gate
voltage in sample 1 in $B=13$~T. Also shown by a dashed line is the
geometric capacitance $C_0$. (b)~The chemical potential as a function
of $\nu$ obtained by integrating the magnetocapacitance; see text. The
zero level corresponds to $\nu=1/2$.}
\end{figure}

A magnetocapacitance trace $C$ as a function of gate voltage $V_g$ is
displayed in Fig.~\ref{fig1}(a) for a magnetic field of 13~T. Narrow
minima in $C$ accompanied at their edges by maxima are seen at
filling factor $\nu\equiv n_shc/eB=1/3$, 2/5, 3/5, 2/7, 3/7, and 4/7.
Near the filling factor $\nu=1/2$, the capacitance $C$ in the range
of magnetic fields studied reaches its high-field value determined by
the geometric capacitance $C_0$ (dashed line). We have verified that
the obtained $C_0$ corresponds to the value calculated using
Eq.~(\ref{C}) from the zero-field capacitance and the density of
states $m/\pi\hbar^2$ (where $m=0.067m_e$ and $m_e$ is the free
electron mass). The chemical potential jump $\Delta\mu$ for electrons
at fractional filling factor can be determined by integrating the
magnetocapacitance over the dip (for more details, see
Ref.~\cite{khrapai07}):
\begin{equation}\Delta\mu=\frac{e}{C_0}\int_{\text{dip}}(C_0-C)dV_g.\label{Delta}\end{equation}

\begin{figure}
\scalebox{0.4}{\includegraphics[clip]{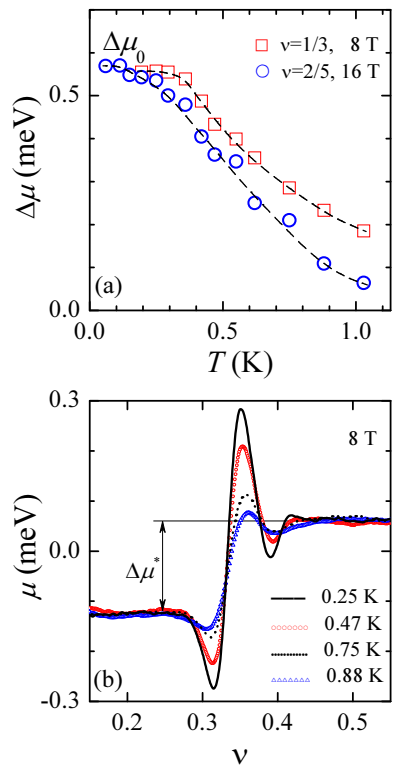}}
\caption{\label{fig2} (a)~Temperature dependence of the fractional
gap in sample 1 for $\nu=1/3$ at $B=8$~T and for $\nu=2/5$ at
$B=16$~T. The saturated low-temperature value $\Delta\mu_0$ is
indicated. The dashed lines are guides to the eye. (b)~The chemical
potential as a function of $\nu$ at different temperatures. The zero
level corresponds to $\nu=1/3$.}
\end{figure}

\begin{figure}
\scalebox{0.42}{\includegraphics[clip]{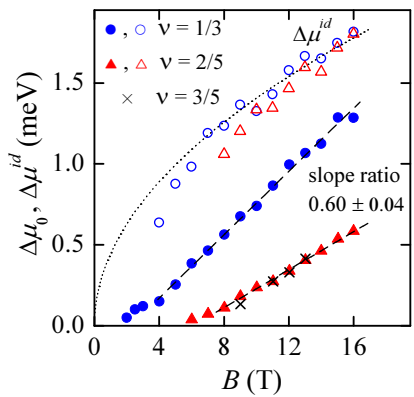}}
\caption{\label{fig3} The fractional gap $\Delta\mu_0$ (filled
symbols and crosses) and jump $\Delta\mu^{id}$ for an ideal 2D
electron system (open symbols) versus $B$ in sample 1. The dashed
lines are linear fits to the high-field data. The slope ratio is equal, within
the experimental uncertainty, to the inverse ratio of the fraction
denominators. The value of $\Delta\mu^{id}$ for $\nu=2/5$ is divided
by the factor of 0.6. The dotted line is a square-root fit to the
high-field data.}
\end{figure}

It is easy to determine the behavior of the chemical potential when
the filling factor traverses the fractional gap, as shown in
Fig.~\ref{fig1}(b). The chemical potential jump corresponds to a cusp
in the dependence of the ground-state energy, $E$, of the 2D electron
system on filling factor \cite{chakraborty00}:
$\Delta\mu(\nu)=\left.dE/dn_s\right|_{\nu+0}-\left.dE/dn_s\right|_{\nu-0}$.
The difference, $\Delta\mu^*$, between the $\mu$ values at $\nu=1/2$
and $\nu=1/4$ is smaller than $\Delta\mu(\nu=1/3)$, as determined by
the maxima in the capacitance near $\nu=1/3$, see below. The values
$\Delta\mu$ for $\nu=2/5$ and $\nu=3/5$ are equal (see also
Fig.~\ref{fig3}), which is already evident from Fig.~\ref{fig1}(a):
the difference $\delta C=C-C_0$ versus $\nu$ is nearly symmetric
about $\nu=1/2$. The electron-hole symmetry in the spin-split Landau
level is characteristic of the high-field data including the
fractional features with $q=3$, 5, and 7.

In Fig.~\ref{fig2}(a), we show the temperature dependence of the gap
for $\nu=1/3$ and $\nu=2/5$. As the temperature is decreased, the
value $\Delta\mu$ increases and in the limit of low temperatures, the
gap saturates and becomes independent of temperature. It is the
saturated low-temperature value $\Delta\mu_0$ that will be studied in
the following as a function of the magnetic field.

In Fig.~\ref{fig2}(b), we compare the temperature dependences of
$\Delta\mu$ and $\Delta\mu^*$. While the jump $\Delta\mu$ decreases
with temperature by more than a factor of two, the value
$\Delta\mu^*$ does not change at all with increasing temperature up
to $T\approx0.9$~K. The qualitatively different behavior of both
values with temperature allows us to conclude that the fractional gap
closes partially (the value of $\Delta\mu_0/\Delta\mu^*$ is about
three for $\nu=1/3$) when the Fermi level lies outside the gap. Since
the gap $\Delta\mu$ decreases with both a deviation of the filling
factor from fractional $\nu$ and temperature in a similar way (i.e.,
approximately linearly), its decay with temperature is likely to be caused
by thermal smearing \cite{dorokhova04}.

In Fig.~\ref{fig3}, we show how the $\nu=1/3$ and $\nu=2/5$ gap
$\Delta\mu_0$ changes with magnetic field. In high magnetic fields,
the data are described by a linear increase of the gap value with $B$,
which is consistent with the results of Ref.~\cite{khrapai07}. We
find that the ratio of the slopes for $\nu=2/5$ and $\nu=1/3$ is the same,
within the experimental uncertainty, as the inverse ratio of the fraction
denominators, equal to 3/5. This shows that the increase of the gap with
magnetic field is determined by the denominator $q$.

\begin{figure}
\scalebox{0.4}{\includegraphics[clip]{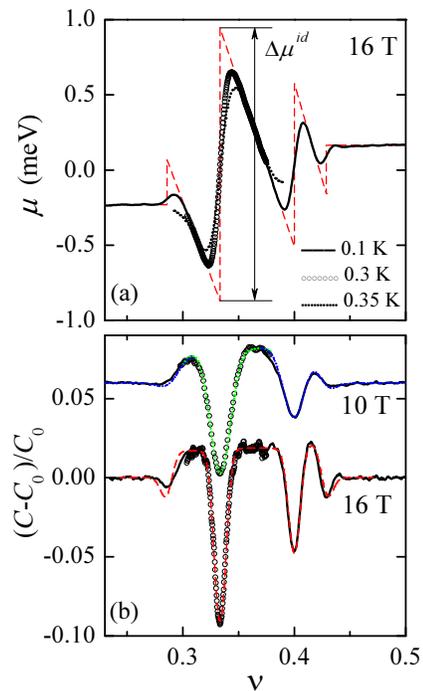}}
\caption{\label{fig4} (a)~Dependence of the chemical potential on
$\nu$ in the low-temperature limit in samples 1 (solid line and
circles) and 2 (dotted line) in $B=16$~T. The dashed red line is a
saw-tooth function obtained by the linear extrapolation to fractional
$\nu$ of the intervals in which the quantity $\mu$ decreases with
filling factor. The zero level corresponds to $\nu=1/3$.
(b)~Magnetocapacitance as a function of $\nu$ in the low-temperature
limit in sample 1 (solid line and circles) at $B=16$~T and $B=10$~T
(vertically shifted for clarity) and the fit based on convolution of
the saw-tooth function with the Gaussian density distribution with
$\sigma(n_s)=2\times10^9$~cm$^{-2}$ (dashed red line) and the fit
based on the experimental $\mu(\nu)$ in $B=16$~T (dotted green and
dash-dotted blue lines); see text.}
\end{figure}

In real samples, the jump of the chemical potential $\Delta\mu_0$ is
washed out due to inhomogeneities in the electron density
distribution. Unlike the capacitance minima at integer $\nu$
\cite{dolgopolov97}, the fractional minima are surrounded by the
relatively big maxima in $C$, which leads to a noticeable reduction
of the disorder-broadened jumps of $\mu$ with respect to those in an
ideal 2D electron system. An attempt to extract values of fractional
gap corresponding to the clean limit was made in
Ref.~\cite{eisenstein92}, based on the theoretical behavior of the
chemical potential around fractional $\nu$ taking into account the
electron-electron interactions. The results obtained do not confirm
the expected square-root dependence of the gap on magnetic field and,
therefore, one should treat electron-electron interactions in the
fractional quantum Hall effect in a less straightforward way.

In Fig.~\ref{fig4}(a), we compare the dependence $\mu(\nu)$ for the
highest $B=16$~T with that in another sample made of the same wafer
but having a considerably higher level of inhomogeneities in the
electron density distribution, as signaled by more broadened and less
deep capacitance minima. Outside the $\nu=1/3$ gap, the decrease of
the chemical potential with filling factor, which corresponds to the
gap closing, can be described by a linear dependence, the same for
both samples. Extrapolation of this linear behavior to $\nu=1/3$ yields
an estimate of the jump $\Delta\mu^{id}$ for an ideal 2D electron
system.

The functional dependence of the gap $\Delta\mu^{id}$ on magnetic
field can be determined under the assumption that the shape of the
oscillations $\mu^{id}(\nu)$ does not depend explicitly on $B$ and
their amplitude is a function of the magnetic field: behavior that is
consistent with theoretical considerations discussed in detail in
Ref.~\cite{eisenstein92}. For the Gaussian density distribution with
constant width $\sigma(n_s)$, the oscillations $\mu^{id}(\nu)$ are
progressively washed out with decreasing magnetic field. Since
$\mu^{id}(\nu)$ near the fractional $\nu$ is in fact unknown, we use
the least-broadened experimental $\mu(\nu)$ in the highest
$B=16$~T for convolution in lower magnetic fields. By tuning the
width $\sigma(\nu)=\sigma(n_s)hc/eB$ and varying the oscillation
amplitude, we attain good fits to the measured $C(\nu)$ around
$\nu=1/3$ and $\nu=2/5$ at $B<16$~T (Fig.~\ref{fig4}(b)) and, thus,
extract the behavior of $\Delta\mu^{id}$ with $B$, as shown in
Fig.~\ref{fig3}. The procedure is applicable in high magnetic fields,
whereas in low $B$ the quality of the fits is poor, making the
determination of $\Delta\mu^{id}(B)$ impossible. Note that the
absolute value of $\Delta\mu^{id}$ in Fig.~\ref{fig3} is an estimate
obtained using linear extrapolations of the data (the dashed line in
Fig.~\ref{fig4}(a)); such saw-tooth oscillations $\mu^{id}(\nu)$ with
zero-width jumps at fractional $\nu$, convolved with the Gaussian
density distribution with $\sigma(n_s)=2\times10^9$~cm$^{-2}$
provide a high-quality fit to the experimental magnetocapacitance,
shown in Fig.~\ref{fig4}(b).

As seen from Fig.~\ref{fig3}, the $\Delta\mu^{id}(B)$ dependence
approaches, ignoring the numerical factor, the theoretical
square-root behavior of the fractional gap with $B$ in the high-field
limit where the ratio of the jumps $\Delta\mu^{id}$ for $\nu=2/5$ and
$\nu=1/3$ is approximately equal to 3/5. In other words, our results
suggest that in the highest accessible magnetic fields, the gap
$\Delta\mu^{id}$ is proportional to $q^{-1}B^{1/2}$. Thus, we reach the
limit of very high magnetic fields where the expected square-root
behavior prevails (in homogeneous samples), while discrepancies
between experiment and theory may remain for lower $B$.

In summary, we have studied the variation of the chemical potential
in the 2D electron system in GaAs/AlGaAs single heterojunctions with
changing filling factor around $\nu=1/3$ and $\nu=2/5$ and determined
the fractional gap in the limit of low temperatures. In high magnetic
fields, both gaps increase linearly with magnetic field with slopes
proportional to the inverse fraction denominator. As inferred from the
qualitatively different behavior with temperature of the jump $\Delta\mu$
and the difference of the $\mu$ values at $\nu=1/2$ and $\nu=1/4$, the
fractional gap closes partially when the Fermi level lies outside the gap.
Unlike the spin gap \cite{dolgopolov97}, the fractional gap decays due to
temperature and disorder smearing. Using an empirical analysis of the
data, we allow for the effect of sample inhomogeneities and extract
the jump $\Delta\mu^{id}(B)$ for an ideal 2D electron system. The
results suggest that in the highest accessible magnetic fields, the
fractional gap in the clean limit is inversely proportional to the
denominator and follows the square-root behavior with magnetic field.

We gratefully acknowledge discussions with S.~V. Kravchenko and
M.~P. Sarachik. We would like to thank J.~P. Kotthaus for an opportunity
to use the clean room facilities at LMU Munich and C. Roessler and C.
Paulus for technical assistance. This work was supported by the RFBR,
RAS, and the Programme ``The State Support of Leading Scientific
Schools''. VSK acknowledges the A.~von Humboldt foundation.

\end{document}